\documentclass[%
 letterpaper,
 superscriptaddress,
preprint,
 amsmath,amssymb,
 aps,
]{revtex4-1}

\usepackage{graphicx}
\usepackage{dcolumn}
\usepackage{bm}
\usepackage[mathlines]{lineno}

\usepackage{units}

\newcommand{\minerva}{\mbox{MINERvA}}

\newcommand{\numu}[0]{\nu_{\mu}}
\newcommand{\nuebar}[0]{\overline{\nu}_{e}}

\newcommand{\nue}[0]{\nu_{e}}
\newcommand{\numubar}[0]{\overline{\nu}_{\mu}}

\begin{document}

\preprint{APS/123-QED}

\title{Measurement of Electron Neutrino Quasielastic and Quasielasticlike scattering on Hydrocarbon at {\bf $\langle E_{\nu} \rangle $} =\ 3.6~GeV }
\newcommand{\deceased}{Deceased}

\newcommand{\Rutgers}{Rutgers, The State University of New Jersey, Piscataway, New Jersey 08854, USA}
\newcommand{\Hampton}{Hampton University, Dept. of Physics, Hampton, VA 23668, USA}
\newcommand{\Dortmund}{Institute of Physics, Dortmund University, 44221, Germany }
\newcommand{\Otterbein}{Department of Physics, Otterbein University, 1 South Grove Street, Westerville, OH, 43081 USA}
\newcommand{\JMU}{James Madison University, Harrisonburg, Virginia 22807, USA}
\newcommand{\Florida}{University of Florida, Department of Physics, Gainesville, FL 32611}
\newcommand{\UCIrvine}{Department of Physics and Astronomy, University of California, Irvine, Irvine, California 92697-4575, USA}
\newcommand{\CBPF}{Centro Brasileiro de Pesquisas F\'{i}sicas, Rua Dr. Xavier Sigaud 150, Urca, Rio de Janeiro, Rio de Janeiro, 22290-180, Brazil}
\newcommand{\PUCP}{Secci\'{o}n F\'{i}sica, Departamento de Ciencias, Pontificia Universidad Cat\'{o}lica del Per\'{u}, Apartado 1761, Lima, Per\'{u}}
\newcommand{\INRM}{Institute for Nuclear Research of the Russian Academy of Sciences, 117312 Moscow, Russia}
\newcommand{\Jlab}{Jefferson Lab, 12000 Jefferson Avenue, Newport News, VA 23606, USA}
\newcommand{\Pittsburgh}{Department of Physics and Astronomy, University of Pittsburgh, Pittsburgh, Pennsylvania 15260, USA}
\newcommand{\Guanajuato}{Campus Le\'{o}n y Campus Guanajuato, Universidad de Guanajuato, Lascurain de Retana No. 5, Colonia Centro, Guanajuato 36000, Guanajuato M\'{e}xico.}
\newcommand{\Athens}{Department of Physics, University of Athens, GR-15771 Athens, Greece}
\newcommand{\Tufts}{Physics Department, Tufts University, Medford, Massachusetts 02155, USA}
\newcommand{\WM}{Department of Physics, College of William \& Mary, Williamsburg, Virginia 23187, USA}
\newcommand{\FNAL}{Fermi National Accelerator Laboratory, Batavia, Illinois 60510, USA}
\newcommand{\Purdue}{Department of Chemistry and Physics, Purdue University Calumet, Hammond, Indiana 46323, USA}
\newcommand{\MCLA}{Massachusetts College of Liberal Arts, 375 Church Street, North Adams, MA 01247}
\newcommand{\UMD}{Department of Physics, University of Minnesota -- Duluth, Duluth, Minnesota 55812, USA}
\newcommand{\Northwestern}{Northwestern University, Evanston, Illinois 60208}
\newcommand{\UNI}{Universidad Nacional de Ingenier\'{i}a, Apartado 31139, Lima, Per\'{u}}
\newcommand{\Rochester}{Department of Physics and Astronomy, University of Rochester, Rochester, New York 14627 USA}
\newcommand{\Austin}{Department of Physics, University of Texas, 1 University Station, Austin, Texas 78712, USA}
\newcommand{\USM}{Departamento de F\'{i}sica, Universidad T\'{e}cnica Federico Santa Mar\'{i}a, Avenida Espa\~{n}a 1680 Casilla 110-V, Valpara\'{i}so, Chile}
\newcommand{\Geneva}{University of Geneva, 1211 Geneva 4, Switzerland}
\newcommand{\Chicago}{Enrico Fermi Institute, University of Chicago, Chicago, IL 60637 USA}
\newcommand{\hired}{}
\newcommand{\OregonState}{Department of Physics, Oregon State University, Corvallis, Oregon 97331, USA}
\newcommand{\bmeThanks}{now at SLAC National Accelerator Laboratory, Stanford, California 94309 USA}
\newcommand{\higueraThanks}{University of Houston, Houston, Texas, 77204, USA}
\newcommand{\damartinezThanks}{Now at Illinois Institute of Technology}
\newcommand{\LazaThanks}{also at Department of Physics, University of Antananarivo, Madagascar}
\newcommand{\twaltonThanks}{now at Fermi National Accelerator Laboratory, Batavia, IL USA 60510}

\author{J.~Wolcott}                       \affiliation{\Rochester}\affiliation{\Tufts}
\author{L.~Aliaga}                        \affiliation{\WM}
\author{O.~Altinok}                       \affiliation{\Tufts}
\author{L.~Bellantoni}                    \affiliation{\FNAL}
\author{A.~Bercellie}                     \affiliation{\Rochester}
\author{M.~Betancourt}                    \affiliation{\FNAL}
\author{A.~Bodek}                         \affiliation{\Rochester}
\author{A.~Bravar}						  \affiliation{\Geneva}
\author{H.~Budd}                          \affiliation{\Rochester}
\author{T.~Cai}                           \affiliation{\Rochester}
\author{M.F.~Carneiro}                    \affiliation{\CBPF}
\author{J.~Chvojka}                       \affiliation{\Rochester}
\author{H.~da~Motta}                      \affiliation{\CBPF}
\author{J.~Devan}                         \affiliation{\WM}
\author{S.A.~Dytman}                      \affiliation{\Pittsburgh}
\author{G.A.~D\'{i}az~}                   \affiliation{\Rochester}  \affiliation{\PUCP}
\author{B.~Eberly}\thanks{\bmeThanks}     \affiliation{\Pittsburgh}
\author{J.~Felix}                         \affiliation{\Guanajuato}
\author{L.~Fields}                        \affiliation{\FNAL}  \affiliation{\Northwestern}
\author{R.~Fine}                          \affiliation{\Rochester}
\author{A.M.~Gago}                        \affiliation{\PUCP}
\author{R.~Galindo}                        \affiliation{\USM}
\author{H.~Gallagher}                     \affiliation{\Tufts}
\author{A.~Ghosh}                         \affiliation{\CBPF}  \affiliation{\Rochester}
\author{T.~Golan}                         \affiliation{\Rochester}  \affiliation{\FNAL}
\author{R.~Gran}                          \affiliation{\UMD}
\author{D.A.~Harris}                      \affiliation{\FNAL}
\author{A.~Higuera}\thanks{\higueraThanks}  \affiliation{\Rochester}  \affiliation{\Guanajuato}
\author{M.~Kiveni}                        \affiliation{\FNAL}
\author{J.~Kleykamp}                      \affiliation{\Rochester}
\author{M.~Kordosky}                      \affiliation{\WM}
\author{T.~Le}                            \affiliation{\Tufts}  \affiliation{\Rutgers}
\author{E.~Maher}                         \affiliation{\MCLA}
\author{S.~Manly}                         \affiliation{\Rochester}
\author{W.A.~Mann}                        \affiliation{\Tufts}
\author{C.M.~Marshall}                    \affiliation{\Rochester}
\author{D.A.~Martinez~Caicedo}\thanks{\damartinezThanks}  \affiliation{\FNAL}
\author{K.S.~McFarland}                   \affiliation{\Rochester}  \affiliation{\FNAL}
\author{C.L.~McGivern}                    \affiliation{\Pittsburgh}
\author{A.M.~McGowan}                     \affiliation{\Rochester}
\author{B.~Messerly}                      \affiliation{\Pittsburgh}
\author{J.~Miller}                        \affiliation{\USM}
\author{A.~Mislivec}                      \affiliation{\Rochester}
\author{J.G.~Morf\'{i}n}                  \affiliation{\FNAL}
\author{J.~Mousseau}                      \affiliation{\Florida}
\author{T.~Muhlbeier}                     \affiliation{\CBPF}
\author{D.~Naples}                        \affiliation{\Pittsburgh}
\author{J.K.~Nelson}                      \affiliation{\WM}
\author{A.~Norrick}                       \affiliation{\WM}
\author{J.~Osta}                          \affiliation{\FNAL}
\author{V.~Paolone}                       \affiliation{\Pittsburgh}
\author{J.~Park}                          \affiliation{\Rochester}
\author{C.E.~Patrick}                     \affiliation{\Northwestern}
\author{G.N.~Perdue}                      \affiliation{\FNAL}  \affiliation{\Rochester}
\author{L.~Rakotondravohitra}\thanks{\LazaThanks}  \affiliation{\FNAL}
\author{R.D.~Ransome}                     \affiliation{\Rutgers}
\author{H.~Ray}                           \affiliation{\Florida}
\author{L.~Ren}                           \affiliation{\Pittsburgh}
\author{D.~Rimal}                         \affiliation{\Florida}
\author{P.A.~Rodrigues}					  \affiliation{\Rochester}
\author{D.~Ruterbories}                   \affiliation{\Rochester}
\author{G.~Salazar}                       \affiliation{\UNI}
\author{H.~Schellman}                     \affiliation{\OregonState}  \affiliation{\Northwestern}
\author{D.W.~Schmitz}                     \affiliation{\Chicago}  \affiliation{\FNAL}
\author{C.J.~Solano~Salinas}              \affiliation{\UNI}
\author{N.~Tagg}                          \affiliation{\Otterbein}
\author{B.G.~Tice}                        \affiliation{\Rutgers}
\author{E.~Valencia}                      \affiliation{\Guanajuato}
\author{T.~Walton}\thanks{\twaltonThanks}  \affiliation{\Hampton}

\author{M.~Wospakrik}                      \affiliation{\Florida}
\author{G.~Zavala}                        \affiliation{\Guanajuato}\thanks{\deceased}
\author{A.~Zegarra}                       \affiliation{\UNI}
\author{D.~Zhang}                         \affiliation{\WM}
\author{B.P.~Ziemer}                       \affiliation{\UCIrvine}

\collaboration{The \minerva\ Collaboration}\ \noaffiliation
\date{\today}


\begin{abstract}

The first direct measurement of electron-neutrino quasielastic and quasielastic-like scattering on hydrocarbon in the few-GeV region of incident neutrino energy has been carried out using the MINERvA detector in the NuMI beam at Fermilab. The
flux-integrated differential cross sections in electron production angle, electron energy and $Q^{2}$ are presented. The ratio of the quasielastic, flux-integrated differential cross section in $Q^{2}$ for $\nue$ with that of similarly-selected $\numu$-induced events from the same exposure is used to probe assumptions that underpin conventional treatments of charged-current $\nue$ interactions used by long-baseline neutrino oscillation experiments.   The data are found to be consistent with lepton universality and are well-described by the predictions of the neutrino event generator GENIE.

\begin{description}
\item[PACS numbers]13.15.+g,25.30.Pt
\end{description}
\end{abstract}

\maketitle


\section{\label{Intro}Introduction}


Current and future neutrino oscillation experiments hope to measure CP violation in the neutrino sector by making precise measurements of $\nue$($\nuebar$) appearance in predominantly $\numu$($\numubar$) beams. 
These experiments (such as NOvA\cite{NOvATDR}, T2K\cite{T2KNIM}, and DUNE\cite{dune})  consist of large detectors of heavy nuclei (e.g., carbon, oxygen, argon) to maximize the rate of neutrino interactions.  They examine the energy distribution of interacting neutrinos and compare the observed spectrum with the predictions based on different oscillation hypotheses.  Correct prediction of the observed energy spectrum for $\nue$ interactions requires an accurate model of the interaction rates, particle content, multiplicity and outgoing particle kinematics. In other words, there is a need for precise $\nue$ cross sections on the appropriate detector materials. 

The relatively small components of $\nue$ and $\nuebar$ flux in neutrino beams coupled with significant backgrounds arising from the dominant $\numu$ interactions have led to a paucity of $\nue$ and $\nuebar$ measurements in this energy range (0.5 to a few GeV).  Gargamelle\cite{Gargamellenue} and T2K\cite{T2Knue} have published $\nue$ inclusive cross-section measurements at these energies, but small statistics and the inclusive nature of both of these measurements limit their usefulness for model comparisons and as a basis for tuning simulations. Therefore,
most simulations, such as those used in oscillation experiments,  begin by tuning to high-precision $\numu$($\numubar$) cross-section data and apply corrections such as those discussed in Ref. \cite{DayMcF} to obtain a prediction for the $\nu_{e}$($\nuebar$) cross section. 

This Letter reports measurements of $\nue$ and $\nuebar$ charged-current quasielastic (CCQE) interactions ($\nu_{e} n \rightarrow e^{-} p$ and $\bar{\nu}_{e} p \rightarrow e^{+} n$) on nucleons in a hydrocarbon target at an average $\nue$ energy of 3.6~GeV.  Quasielastic scattering is a two-body process that is of particular importance in neutrino physics since it is the dominant reaction near 1 GeV, which is a critical energy region for accelerator-based long-baseline oscillation experiments.  Though the incoming neutrino has an unknown energy and the final-state nucleon may not be detected, knowledge of the incoming neutrino direction and the outgoing lepton momentum vector, along with the assumption that the initial-state nucleon is at rest, are sufficient to constrain the kinematics.  Thus the assumption that quasielastic scattering takes place on free, stationary nucleons is often used to extract an estimate of the neutrino energy and the square of the four-momentum transferred to the nucleus ($E_{\nu}^\mathrm{QE}$ and $Q^{2}_{\mathrm{QE}}$, respectively).  However, hadrons exiting the nucleus after the interaction can reinteract and change identity or eject other hadrons\cite{GiBUUFSI}, and the complex interactions within the initial nuclear environment can deform the inferred kinematics or cause multiple nucleons to be ejected by a single interaction\cite{Martinicorr,Nievescorr}. Thus, true quasielastic events cannot be reliably isolated experimentally.  As an alternative, this analysis defines ``CCQE-like'' events to be the signal. These are events having a prompt electron or positron from the primary vertex plus any number of nucleons but devoid of any other hadrons or associated $\gamma$-ray conversions. Both $\nue$- and $\nuebar$-induced CCQE-like events are included since the final-state $e^{\pm}$ cannot be distinguished in \minerva 's unmagnetized tracking volume.
 The $\nuebar$ have a significantly smaller flux and cross section relative to the $\nue$, though there is a small analysis selection bias favoring $\nuebar$ over $\nue$.  According to the simulation, the $\nuebar$-induced events comprise 8.9\% of the selected sample of $\nue$ and $\nuebar$ interactions. In this Letter, the $\nuebar$ (positron) content is included when referring to the signal.

The relatively high statistics in the $\minerva$ data set allows for flux-integrated differential cross-section measurements for the $\nue$ quasielasticlike process as well as a comparison of the $\nue$ and $\numu$ quasielastic cross sections as a function of $Q^{2}_{\mathrm{QE}}$.  
These measurements are useful for neutrino oscillation experiments seeking to quantify their understanding of the expected  $\nue$ energy distribution. Notably, the target medium for this analysis (hydrocarbon) is nearly identical to that used in NOvA and the T2K near detector, and the neutrino energy range of this analysis overlaps that of NOvA and DUNE.

\section{\label{ExptData}The \minerva\ experiment}

\minerva\ records interactions of neutrinos produced in the NuMI
beam line\cite{NUMIref}. In NuMI, a beam of 120-GeV protons
strikes a graphite target and produces charged mesons which are focused by two magnetic horns into a 675-m
helium-filled decay pipe where most of the charged mesons decay producing neutrinos.  For the data used in this analysis, the horns focused positive mesons, resulting in a beam enriched in neutrinos with a most probable neutrino energy of $3.1$~GeV.
This analysis uses data taken between March 2010 and April 2012 with $3.49\times 10^{20}$ protons on target (POT).

\begin{figure}[h]
			\centering
			\includegraphics[width=0.5\textwidth]{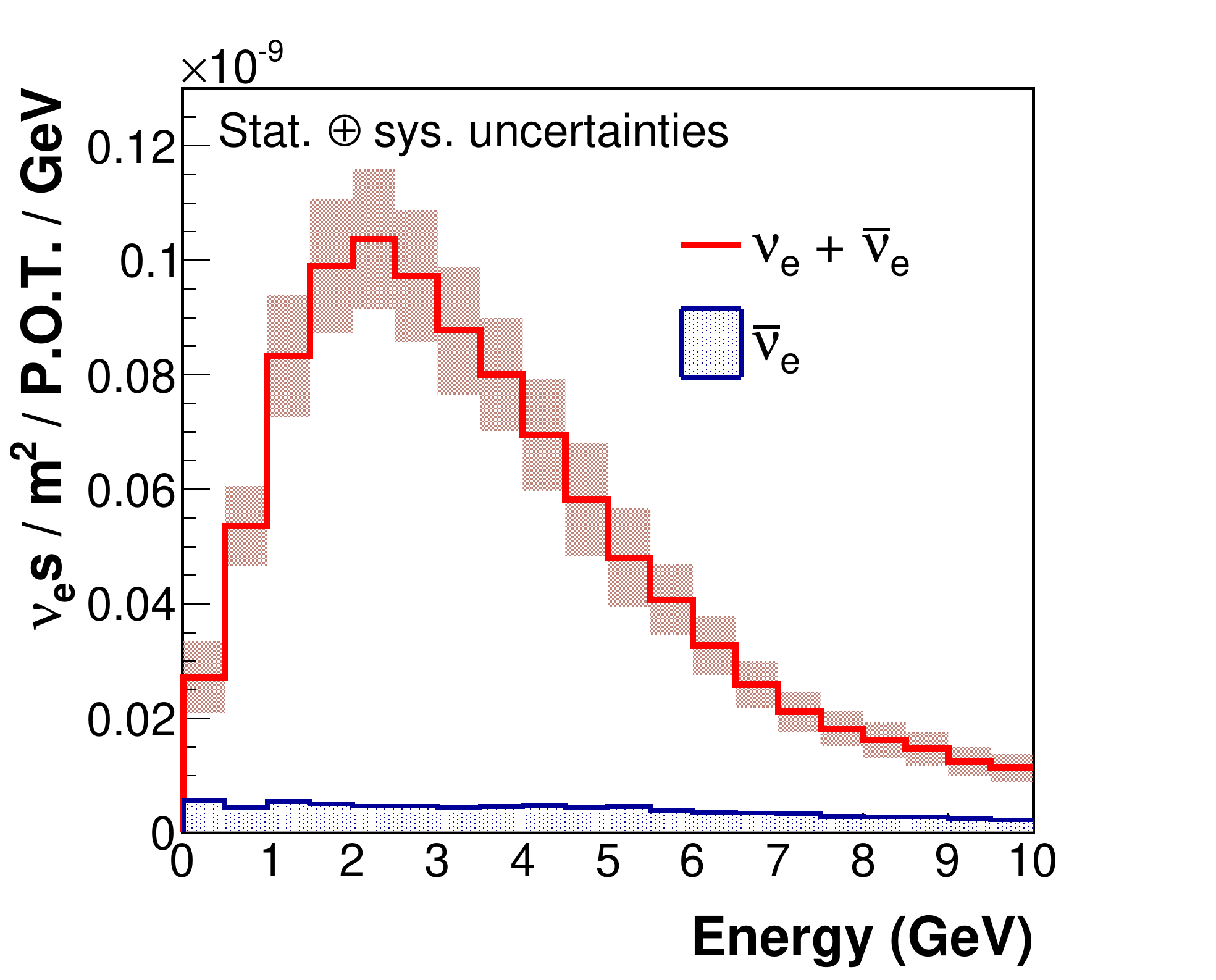}
			\caption{The $\nue + \nuebar$ flux  as a function of neutrino energy from the beam simulation for the data used in this analysis. The  $\nuebar$ flux is shown separately to emphasize the dominance of $\nue$ in the sum. }
			\label{fig:flux}
		\end{figure}

The neutrino beam is simulated by a GEANT4-based
model\cite{Agostinelli,Allison} constrained to
reproduce hadron production measurements\cite{Alt,Barton,Baatar,denisov,carroll,Abgrall,Tinti,Lebedev,Allaby}.
Hadronic interactions not constrained by the 
external hadron production measurements
are predicted
using the Fritiof Precompound (FTFP) hadron shower model\cite{FTFP}.  The uncertainty on the prediction of
the neutrino flux depends upon the precision in these hadron production
measurements, uncertainties in the beam line focusing system and
alignment\cite{Pavlovic}, and comparisons between different
hadron production models in regions not covered by the external data.
Recently,  an \textit{in situ} \minerva\ measurement of purely leptonic $\nu-e$ elastic scattering from atomic electrons\cite{JaewonThesis} became available and can be used to provide a data-based constraint for the flux estimate by comparing the precisely predicted rate for this process with what is observed.  The calculated $\nue + \nuebar$ flux for the analysis in this Letter, which includes the application of the $\nu$-$e$ constraint, is shown in Fig.~\ref{fig:flux} and is provided in tabular form in the ancillary material.

The \minerva\ detector
consists of a core of scintillator strips surrounded by electromagnetic and hadronic calorimeters on the
sides and downstream end of the detector. The 
target and tracking region for this analysis is 95\% CH and 5\% other materials by
weight.  The triangular $3.4\times1.7$~cm$^2$ strips are approximately perpendicular to the beam axis and are arranged in hexagonal planes of three orientations, enabling stereoscopic reconstruction of the neutrino interaction vertex and outgoing charged tracks. The downstream electromagnetic calorimeter (ECAL) is identical to the tracking region except for the addition of a 0.2-cm (0.35 radiation lengths) lead sheet in front of every two planes of scintillator. 

\minerva\ is located 2 m upstream of the MINOS near detector,
a magnetized iron spectrometer\cite{Michael:2008bc}, which is  used
to reconstruct the momentum and charge of $\mu^\pm$.
The \minerva\ detector's response is simulated by a tuned GEANT4-based\cite{Agostinelli,Allison} program.
The energy scale of the detector is set by ensuring that both the photostatistics and the reconstructed energy deposited by
momentum-analyzed beam-related muons traversing the detector agree in the data and simulation.
The calorimetric constants used to reconstruct the energy of
electromagnetic showers, including corrections for passive material\cite{MINERvA_NIM} and algorithm-specific tuning, are determined from the simulation.  Detailed descriptions of the \minerva\ detector configuration, calibrations, and performance can be found in Refs. \cite{MINERvA_NIM,MINERVA_TB,MINERVA_DAQ}.

Neutrino interactions are simulated using the GENIE 2.6.2 event generator\cite{GENIE262}. The simulation is used for efficiency corrections, unfolding, and background estimation. Weak interaction [vector minus axial-vector (V-A)] phenomenology is used for quasielastic interactions\cite{LlewellynSmith:1971zm} in the simulation, with axial mass $M_{A}$=0.99~GeV and a relativistic Fermi gas nuclear model.  The modeled charged-current cross sections differ for $\nue$ and $\numu$ only in the lepton mass, which appears in kinematic factors in the differential cross-section expressions.

\section{\label{reconanal}Event reconstruction and analysis}

Events selected for this analysis are required to originate from a 5.57-ton fiducial volume in the central scintillator region of \minerva.  The energy depositions in the scintillator strips (hits) are first grouped
in time and then spatially grouped into clusters of energy
 in each scintillator plane.  Clusters with energy $>1 ~\rm{MeV}$ are matched among the three views to create tracks.   The hits in each scintillator strip are recorded with 3.0-ns timing resolution, allowing separation of  multiple interactions within a single beam spill.  Candidate events are 
created from tracks whose most upstream energy deposition is in the fiducial volume and which do not exit the back of the detector, as such highly penetrating tracks are overwhelmingly muons. All  tracks passing the criteria above are tested as $e^{\pm}$ candidates.  Hits are considered if they fall within a region that consists of the union of two volumes:  a cylinder of  radius \unit[50]{mm} extending from the event vertex along the track direction and a $7.5^{\circ}$ cone with an apex at the event vertex (origin of track) and a symmetry axis along the track direction.  Hits are associated with the cone as it extends through the scintillator tracker and ECAL; the collection of hits ceases when a gap of three radiation lengths is encountered that is devoid of hits.
The hits in this cone ``object'' are examined using a multivariate particle identification (PID) algorithm.  This technique combines details of the energy deposition pattern both longitudinally (mean $dE/dx$ and the fraction of energy at the downstream end of cone) and transverse to the axis of the cone (mean shower width) using a $k$-nearest-neighbors algorithm\cite{knnref}.
For those candidate events deemed consistent with an electromagnetic cascade, electrons and positrons are separated from photons by demanding the energy deposition near the upstream end of the cone be consistent with a single track rather than the two particles expected from photon conversion to $e^{+}e^{-}$.
The discriminant used for this separation is the minimum energy in a sliding 100-mm window along the axis of the cone, in 20-mm steps, from the event origin up to 500~mm (about 1.2 radiation lengths).  This technique reduces the possibility of bias introduced by nuclear activity near the interaction point\cite{numuPRL}. Cone objects surviving to this point are considered to be electron (or positron) candidates.

The next stage of the analysis requires the topology of the event to be consistent with $\nue$ CCQE-like.
Events containing tracks consistent with charged pions or muons or events with electromagnetic activity outside of the electron candidate cone object (such as might be expected in the presence of a $\pi^{0}$ decay) are removed by a cut on the ``extra energy ratio" variable $\Psi$.  This quantity represents the relative amount of energy outside the electron candidate cone  to that inside the electron candidate cone.  Hits within a sphere of 30-cm radius about the interaction vertex are ignored when calculating $\Psi\ $ to reduce the contribution from low-energy nucleons which are  potentially not well simulated \cite{numuPRL}.
Events at large $\Psi$ are removed from the sample.  The cut in $\Psi$ is a function of the total visible energy of the event and was tuned using simulated events.  In addition to the $\Psi$ cut,  Michel electron candidates from the $\pi \rightarrow \mu \rightarrow e$ decay chain are rejected via timing and their spatial proximity to track ends.

Finally, events are retained in the sample only if they have a reconstructed electron energy $E_{e}$ greater than 0.5 GeV and a reconstructed neutrino energy $E_{\nu}^\mathrm{QE}$ less than \mbox{10 GeV}.  The lower bound excludes a region where the expected flux of $\nue$ and $\nuebar$ is small and the backgrounds are high. The upper bound eliminates events in the region of large flux uncertainty.

 		\begin{figure}[h]
			\centering
			\includegraphics[width=0.5\textwidth]{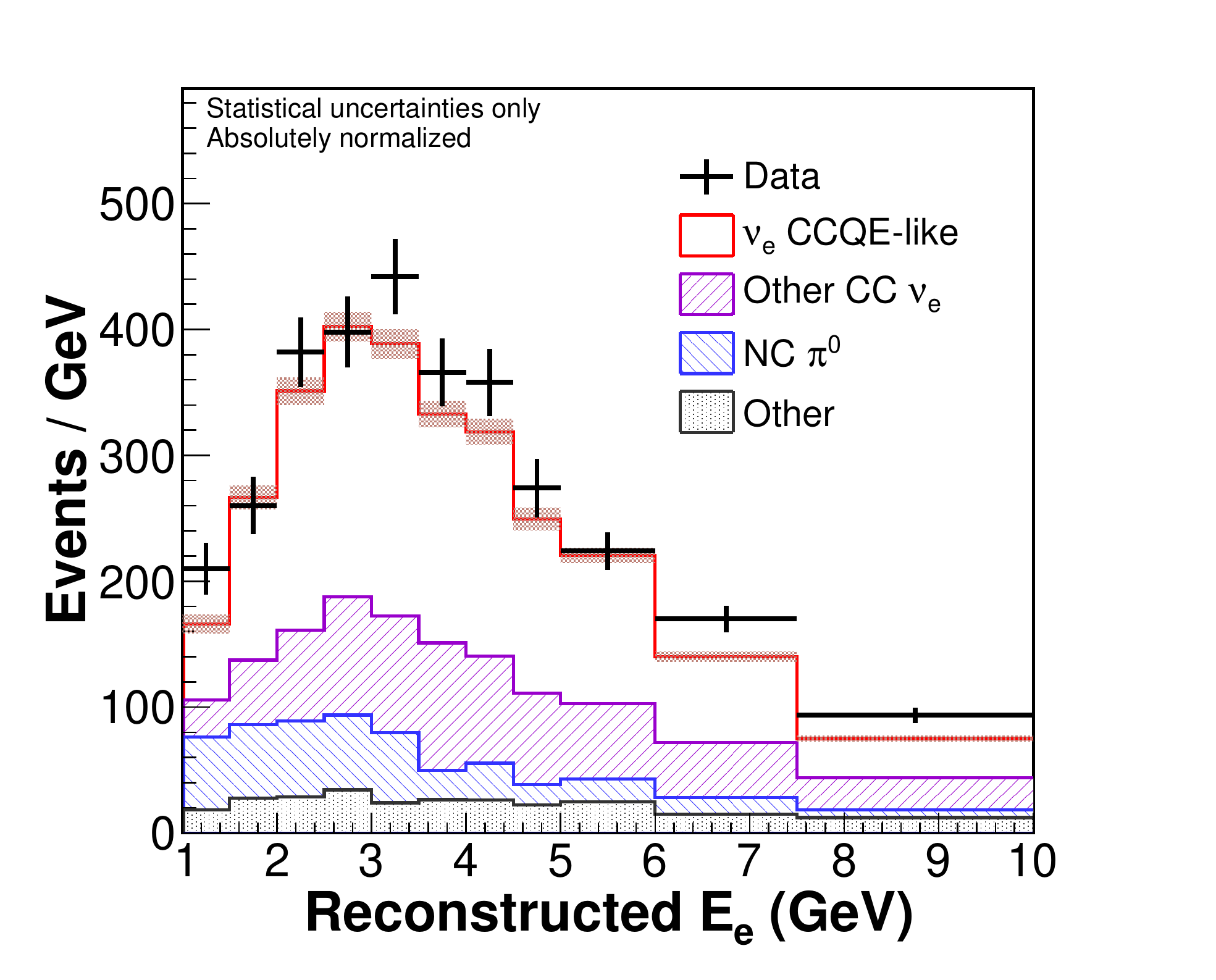}
			\caption{The reconstructed electron energy distribution after all selection cuts and after constraining the backgrounds using sidebands in the data.  The errors shown on the data are statistical only.}
			\label{fig:selectedsample}
		\end{figure}


The reconstructed electron energy distribution of the 2105 selected $\nue$ CCQE-like candidates is shown in Fig.~\ref{fig:selectedsample} for both the data and the simulated event samples.  The simulated sample is broken down by process according to the GENIE event generator and is 52\% pure signal events.  The primary source of background in the selected sample arises from $\nue$-induced non-CCQE-like events.  The second largest background comes from incoherent neutral current (NC) $\pi^{0}$ production.
Coherent $\pi^{0}$ production and neutrino-electron elastic scattering
also contribute to the final sample.

The sizes of the backgrounds in Fig.~\ref{fig:selectedsample} are constrained by two sideband samples.  The first sideband consists of events at larger $\Psi$, which is enriched in inelastic backgrounds from $\nue$ interactions and incoherent events containing $\pi^{0}$.  The other sideband, dominated by $\nue$ CC inelastic events, consists of events with Michel electron candidates (where the Michel electron was typically produced via the decay chain of a charged pion).
The normalizations  of the $\nue$ inelastic and incoherent  $\pi^{0}$ backgrounds are varied in order to find the best overall fit of simulation to the data in the reconstructed electron angle and reconstructed electron energy distributions in each sideband sample.  Since, according to the simulation, the sideband in $\Psi$ contains some signal events, the procedure is iterative.  The background scale fit is done and the signal is extracted and used as a constraint for a new background scale fit.  This is done until the background scale factors stabilize (two iterations).  After this procedure, the fitted scale factor for the normalization for the $\nue$ inelastic category is found to be $0.89 \pm 0.08$, while that for the incoherent $\pi^{0}$ processes is $1.06 \pm 0.12$. The neutral-current coherent pion production  is scaled down by a factor of 2 for pions with energies below 450 MeV in the simulation to bring the GENIE charged-current coherent charged pion production into agreement with a recent \minerva\ measurement \cite{minervaCCcoh}.
Subsequent to these constraints, the scaled backgrounds in the signal region are subtracted from the data. 

An excess of  photon or $\pi^{0}$-like events in the data relative to the simulation was observed in the distribution of energy deposited in the upstream part of the electron candidate cone, as characterized and described in detail in another paper\cite{excessPRL}.  Models of single photon or $\pi^{0}$ production consistent with the observed excess were evaluated and found to have little effect on the background in the signal region of this analysis. Nevertheless, a $\pi^{0}$ background fitted to the excess is added into the simulation and contributes (negligibly) to the background subtraction.   

The flux-integrated differential cross sections in electron energy $E_{e}$, angle $\theta_{e}$, and four-momentum transfer $Q^{2}_\mathrm{QE}$
are calculated in bins $i$ as a function of sample variable $\xi$, with $\epsilon$ representing signal acceptance, $\Phi$ the flux integrated over the energy range of the measurement (or over the bin $i$, in the case of the total cross section), $T_{n}$ the number of targets (nucleons) in the fiducial region, $\Delta_{i}$ the width of bin $i$, and $U_{ij}$ a matrix, derived from the simulation, correcting for detector smearing between bins $i$ and $j$ in the variable of interest:
		\begin{equation}
			\label{eq:dsigma}
			\left( \frac{d\sigma}{d\xi} \right)_{i} = \frac{1}{\epsilon_{i} \Phi T_{n} \Delta_{i}} \times \sum_{j}{U_{ij} \left(N_{j}^{\mathrm{data}} - N_{j}^{\mathrm{bknd\ pred}}\right)}.
		\end{equation}
$E_{\nu}^\mathrm{QE}$ and $Q^{2}_\mathrm{QE}$ are calculated from the lepton kinematics alone using the  approximation of a stationary target nucleon.  Unfolding to correct for detector effects in the four variables is done using a Bayesian technique\cite{dagostini} with a single iteration.  

The systematic errors considered arise from the primary neutrino interaction model, the flux model, and the detector response to particle activity.  The errors on the flux are determined as discussed earlier.  At the focusing peak, i.e., those neutrinos most relevant for this analysis, the $\nue$ flux arises from muons from pion decays.  
The errors in the primary neutrino interaction model are evaluated via the reweighting of events by varying the underlying model tuning parameters according to their uncertainties.
The parameters varied in this way include  the shape and normalization  for elastic and resonance productions, nuclear model parameters principally affecting the deep inelastic scattering, and parameters which control the strength and behavior of the final-state interactions.  Contributions to the  detector response systematic error were determined by varying the energy scale for electromagnetic  interactions, the parameter used in Birks' law, the photomultiplier tube cross-talk fraction, the Michel electron reconstruction energy scale, and the detector mass. 
The largest systematic errors contributing to the cross-section results presented here are due to the detector response, the interaction model, and the flux model, with each  contributing a fractional uncertainty of less than 10\%.  The overall systematic errors are typically in the 10\%-15\% range, which is sufficiently small for the results presented here to be statistically limited.

\begin{figure*}
\hspace{-0.015\textwidth}
\includegraphics[width=0.48\textwidth]{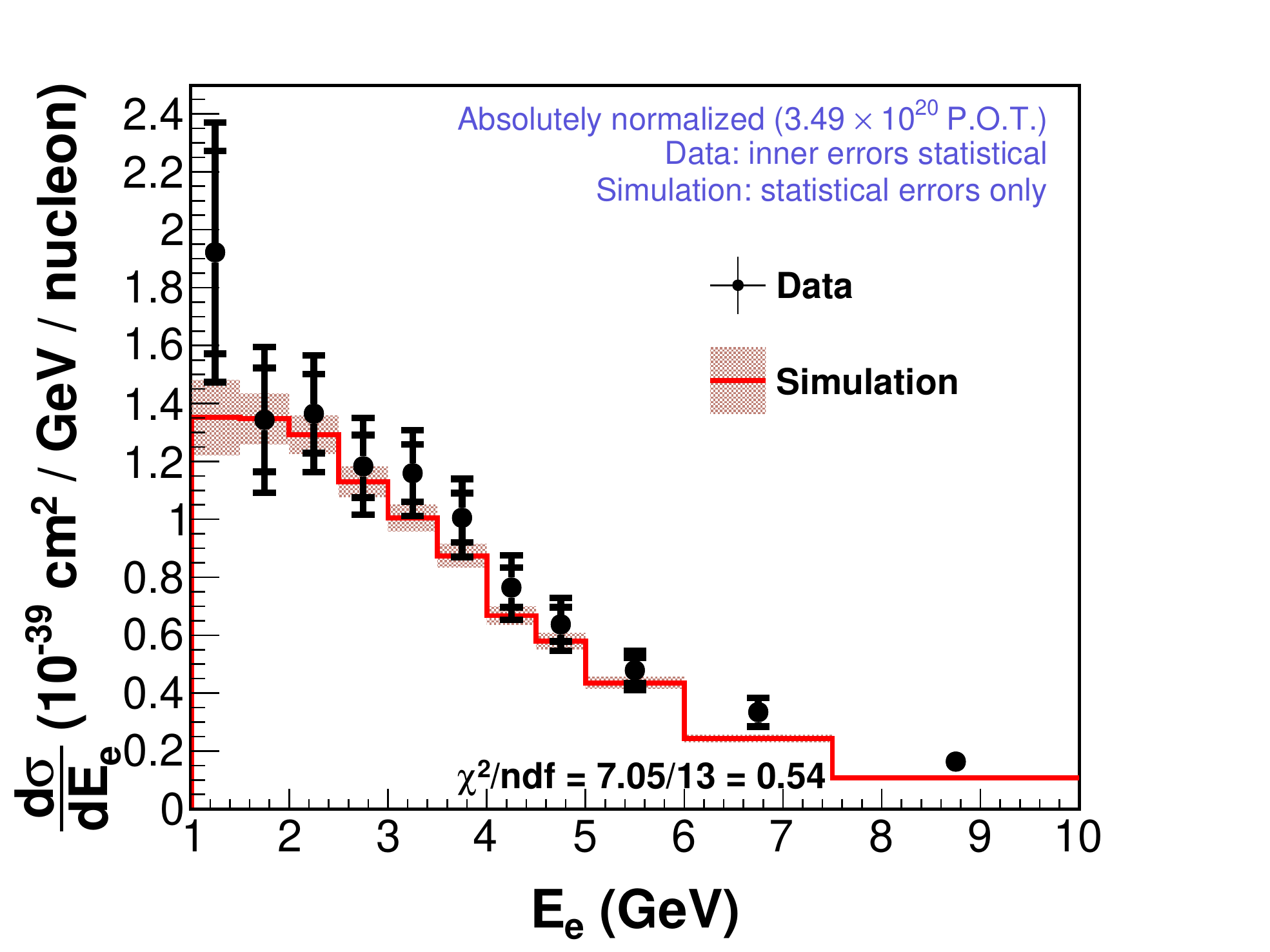}
\hspace{+0.02\textwidth}
\includegraphics[width=0.48\textwidth]{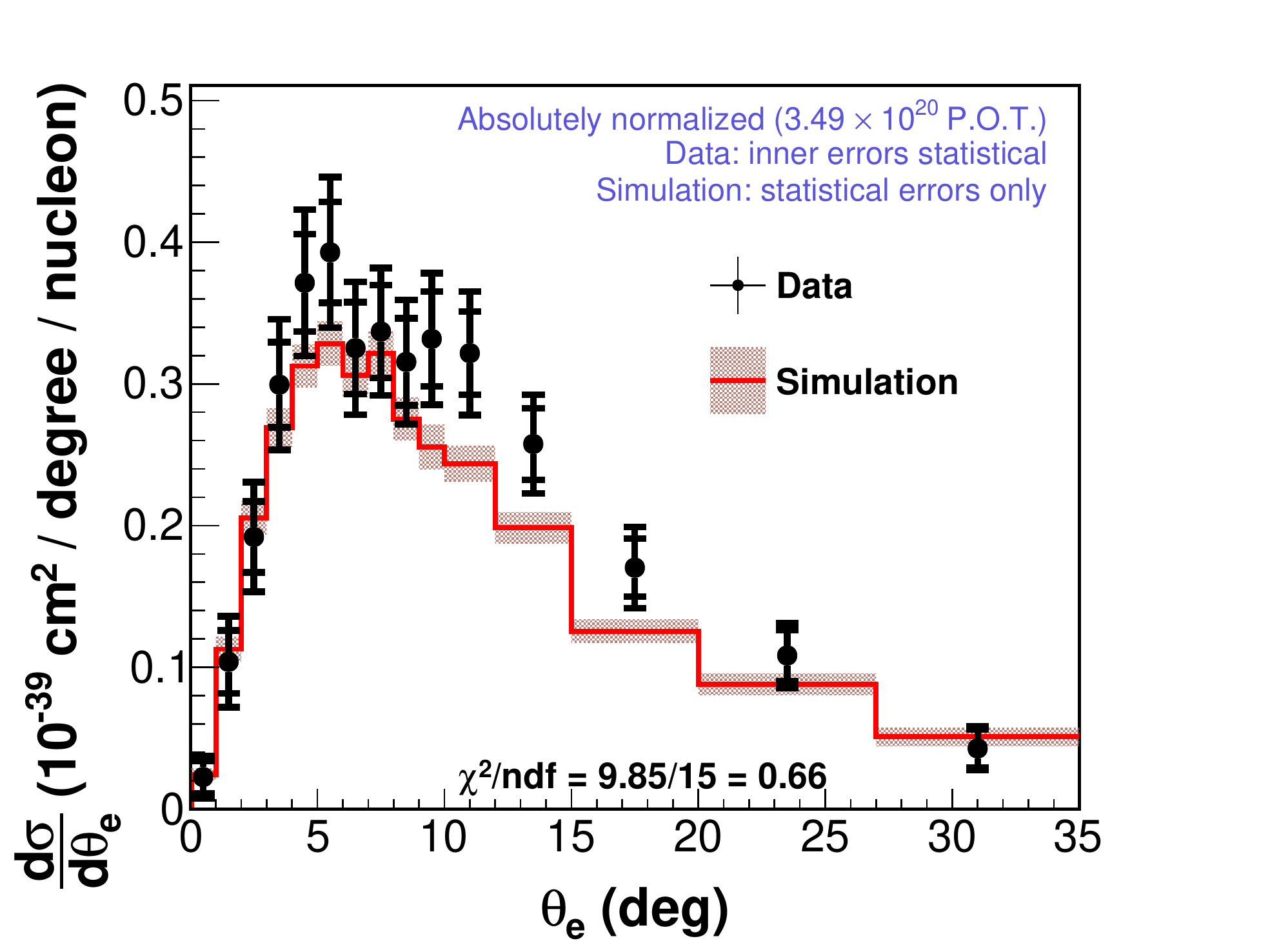}
\caption{Flux-integrated differential $\nue$ CCQE-like cross section versus electron energy (left) and electron angle (right).  
Inner errors are statistical; outer are statistical added in quadrature with systematic.  The band represents the statistical error for the Monte Carlo curve.  }
\label{xsecplots}
\end{figure*}

\begin{figure*}
\hspace{-0.015\textwidth}
\includegraphics[width=0.48\textwidth]{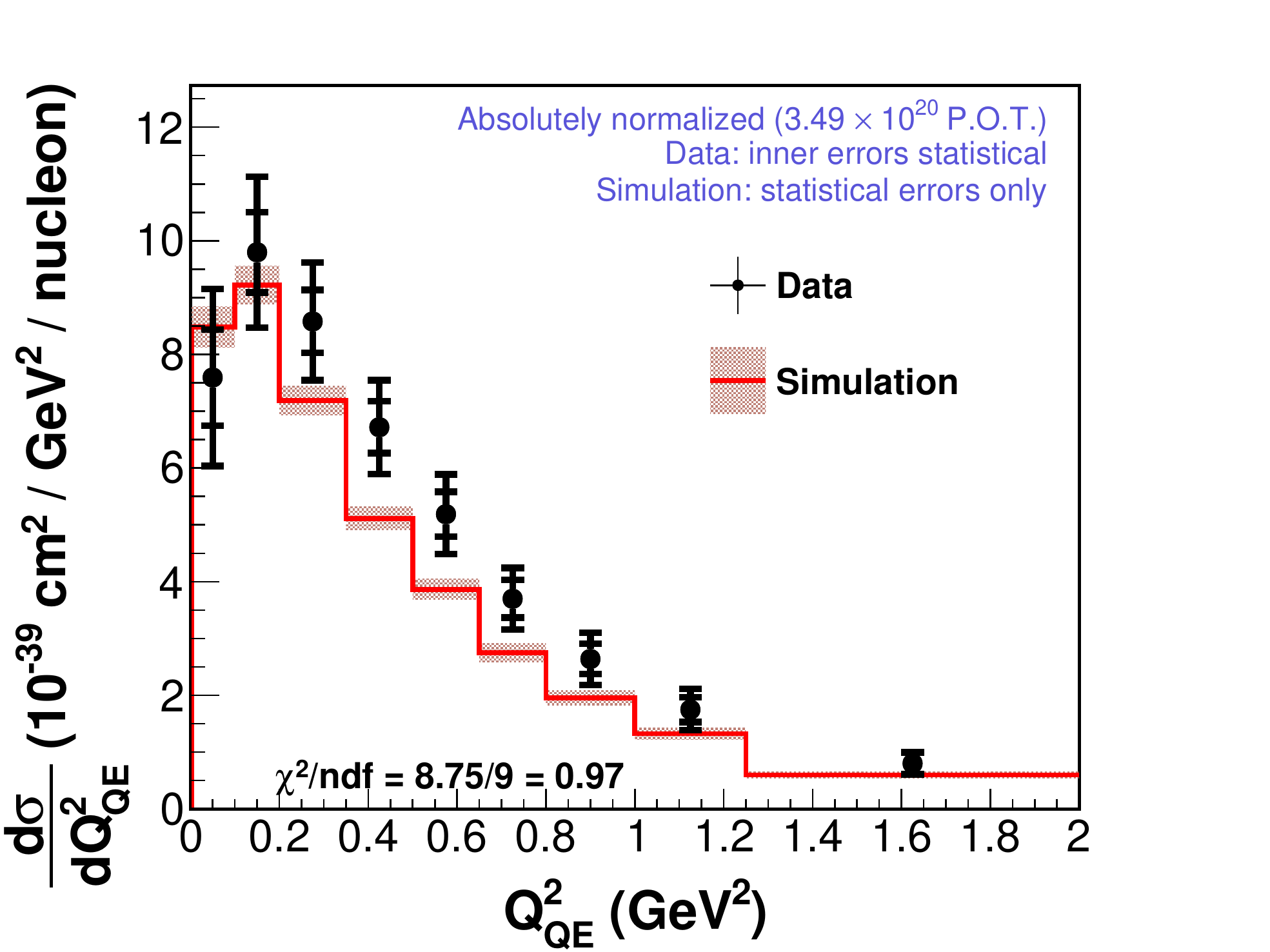}
\hspace{+0.02\textwidth}
\includegraphics[width=0.48\textwidth]{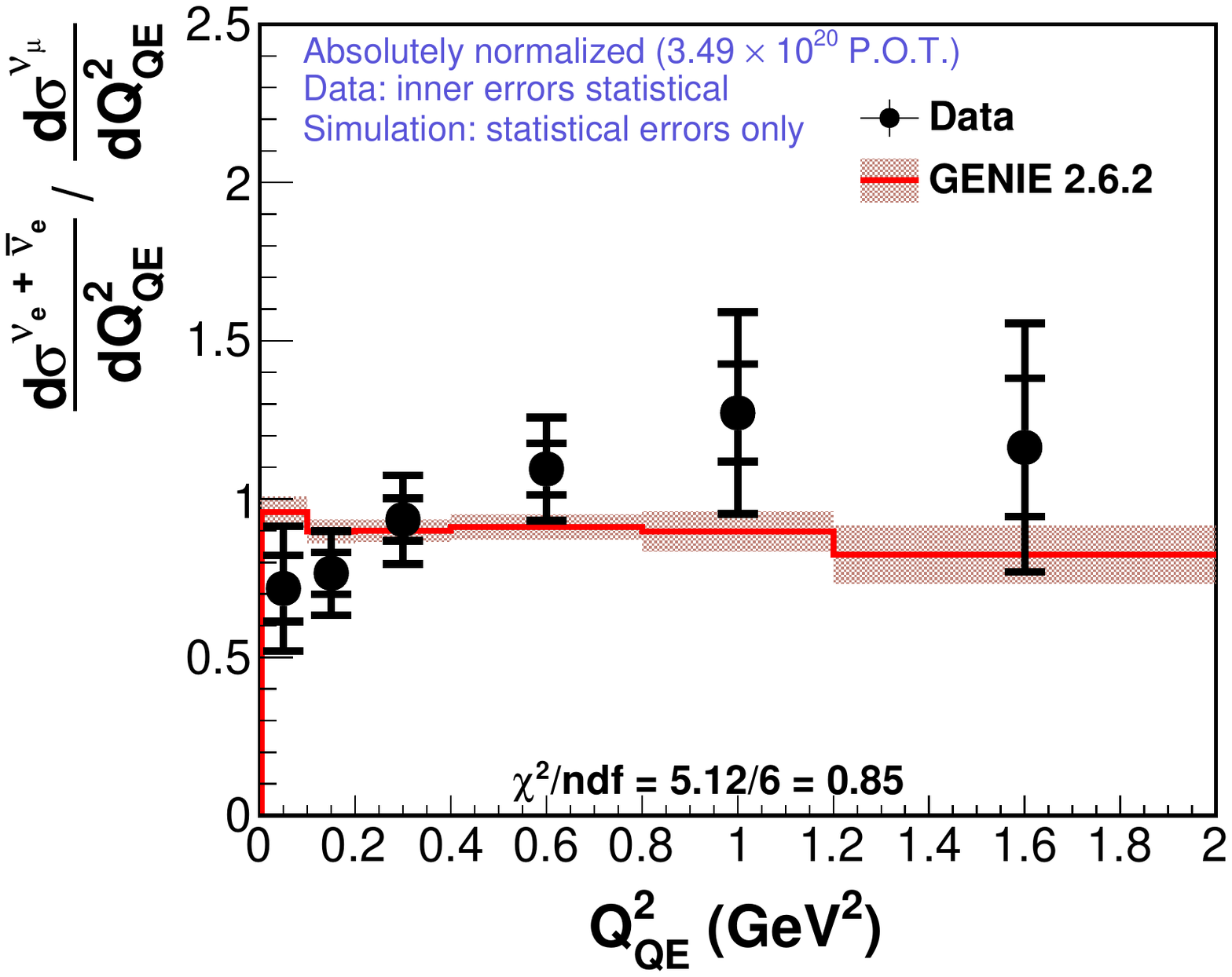}
\caption{The flux-integrated differential $\nue$ CCQE-like cross section versus $Q^{2}_\mathrm{QE}$ (left).  
Inner errors are statistical; outer are statistical added in quadrature with systematic.  On the right is shown the ratio of the \minerva\ $\nue$ CCQE differential cross section as a function of $Q^{2}_\mathrm{QE}$ to the analogous result from \minerva\ for $\numu$\cite{numuPRL}. In both figures, the band represents the statistical error for the Monte Carlo curve.  }
\label{xsecQ2plots}
\end{figure*}

The flux-integrated differential $\nue$ CCQE-like cross sections versus electron energy and angle are given in Fig.~\ref{xsecplots}, for both the data and the POT-normalized Monte Carlo samples.  The analogous distribution in $Q^{2}_\mathrm{QE}$ is given on the left side of Fig.~\ref{xsecQ2plots}.  The measured cross sections and covariances are provided in tabular form in the ancillary material.  The simulation appears to underestimate the width of the electron production angle and exhibit a harder spectrum in $Q^{2}_\mathrm{QE}$. However, these differences are not significant when correlated errors, such as the electromagnetic energy scale, are taken into account.

        In order to compare directly the measured differential cross section for $\nue$ and $\numu$ interactions on carbon as a function of $Q^{2}_{\textrm{QE}}$, an analysis similar to that described in this Letter was performed in terms of a CCQE signal (rather than CCQE-like), as specified by the GENIE event generator, which can be compared directly to previously published \minerva\ results\cite{numuPRL}.  The selection cuts for the $\nue$ events were adjusted slightly to ensure the energy range of included events agreed with that of the $\numu$ analysis.  The ratio of these two results and the corresponding ratio of the Monte Carlo predictions are given on the right in Fig.~\ref{xsecQ2plots}.  The data for the differential cross section for $\nue$ CCQE interactions agree within errors with that for $\numu$ CCQE interactions.  (Some of the uncertainties evaluated in this analysis, such as the electromagnetic energy scale, result in $Q^{2}$-dependent changes to the data distribution shape.  These can cause trends similar to the upward slope in Fig.~\ref{xsecQ2plots}.  When accounting for these correlations, the shape of the data curve is consistent with the shape of the GENIE prediction within 1$\sigma $.)

\section{\label{concl}Conclusions}

This Letter presents the first exclusive measurement of the flux-integrated differential cross section  for $\nue$ CCQE-like interactions and thus provides the first data for directly testing and tuning models of a critical channel for accelerator-based oscillation experiments.  The flux-integrated differential distributions of the cross section in $E_{e}$, $\theta_{e}$ and $Q^{2}_\mathrm{QE}$ agree with the expectation from lepton universality.  A direct comparison, in the same detector, of the differential flux-integrated cross section of $\nue$ CCQE interactions to that for  $\numu$ CCQE interactions as a function of $Q^{2}_\mathrm{QE}$ also shows good agreement.  Collectively, these measurements constitute an important first test of the common assumption made by oscillation experiments that $\numu$ cross-section data can be applied to models of $\nue$ CCQE interactions.

\begin{acknowledgments}

This work was supported by the Fermi National Accelerator Laboratory
under US Department of Energy Contract
No. DE-AC02-07CH11359 which included the \minerva\ construction project.
Construction support was
also granted by the United States National Science Foundation under
Award No. PHY-0619727 and by the University of Rochester. Support for
participating scientists was provided by NSF and DOE (USA), by CAPES
and CNPq (Brazil), by CoNaCyT (Mexico), by CONICYT (Chile), by
CONCYTEC, DGI-PUCP and IDI/IGI-UNI (Peru), by Latin American Center for
Physics (CLAF), and by RAS and the Russian Ministry of Education and Science (Russia).  We
thank the MINOS Collaboration for use of its
near detector data. We acknowledge the dedicated work of the Fermilab staff responsible for the operation and maintenance of the beam line and detector.

\end{acknowledgments}


\end{document}